# Gender Inequalities: Women Researchers Require More Knowledge in Specific and Experimental Topics

Shiqi Tang[1], Dongyi Wang[1], Jianhua Hou[1,*]

[1] School of Information Management, Sun Yat-sen University, P. R. China

**Abstract**

Gender inequalities in science have long been observed globally. Studies have demonstrated it through survey data or published literature, focusing on the interests of subjects or authors; few, however, examined the manifestation of gender inequalities on researchers' knowledge status. This study analyzes the relationship between regional and gender identities, topics, and knowledge status while revealing the female labor division in science and scientific research using online Q&A from researchers. We find that gender inequalities are merged with both regional-specific characteristics and global common patterns. Women's field and topic distribution within fields are influenced by regions, yet the prevalent topics are consistent in all regions. Women are more involved in specific topics, particularly topics about experiments with weaker levels of knowledge and they are of less assistance. To promote inequality in science, the scientific community should pay more attention to reducing the knowledge gap and encourage women to work on unexplored topics and areas.

**Introduction**

The gender gap in the scientific workforce has long been an issue of concern. At the global level, women account for about 30% of researchers, comprising 48.5% for Central Asia, 45.8% for Latin America and the Caribbean, 40.9% for Arab States, 39% for Central and Eastern Europe, 32.9% for North America and Western Europe, 31.1% for Africa, 25% for East Asia and the Pacific and 23.1% for South and West Asia in 2017[1]. Women publish fewer research papers on average than male researchers and are less likely to collaborate internationally[2-5], let alone be editors or in leadership positions within their fields[6-8]. The participation of women also varies by field. Generally, women are under-represented in geosciences, engineering, economics, mathematics, computer science, and physical science (particularly STEM) and over-represented or at parity in the life sciences, psychology, and social sciences[2, 9-11]. Even in specific fields, disparities exist in men's and women's research topic selection. One of the commonly mentioned differences is that male dominance led to little attention to sex differences [12, 13], while females are more likely to produce discoveries about women[14, 15] in life science and medicine. Females are observed in people topics, males in thing topics[16-18] and males in theoretical core topics, and females in peripheral applied topics[19] in STEM fields.

These gaps, from the macro-level of the total women workforce in different regions and fields to the micro-level of their topic selection within research fields, are all manifestations of gender inequality in global science. Such inequality can be expressed from two aspects: the first one is inequality as unequal outcomes, and the second one is inequity as the degree to which these outcomes are a result of impartiality or bias in judgment [20]. We could see inequality for women, including that they are more likely to be under-represented in scientific productivity and impact owing to having less chance to collaborate and unequal outcomes of peer review[21-25], and it is harder for women to gain funding and advance their career due to evaluation and their research topic selection[26-29]. These inequalities may be a result of several implicit biases, including the stereotype of gender that men have more ability, especially in STEM fields[30, 31], men are committed to science careers [32, 33], and women have empathy and hard work while men for genius to gain success[34], as well as general gender bias caused by some social, cultural and economic factors[3], showing that women in the scientific community are still suffering from inequity.

Researchers have already shown that knowledge is distributed unevenly in regions, differentiated according to geography[35], history[36], existing scientific strength[37-39], and economic conditions[40, 41]. As a result, the gender distribution of fields and topics in different regions should be various as a manifestation of science, showing regional-specific characteristics. Besides, what remains a puzzle is that although regions with high and low incomes are at different stages in science development, they may exhibit universal gender inequalities with some commonalities. To reveal the manifestation of common inequalities and the specific characteristics of different fields in various regions, studies should consider gender gaps combined with regional identities.

An increasing number of works examine the relationship between gender identities and their field or topic selection using professional survey data or published scientific literature[12-19]. For example, females are of people-related interests, while males are of thing-related interests at the field level[17, 18]. Yet, a more detailed study on the topic raised the puzzle that women are over-represented in life science, which does not necessarily involve people, but males are represented in density and surgery, which are people-based[42], calling for a new description of the distribution mode. Taken together, these studies show intrinsic characteristics to specific fields or topics that make them more or less appealing to researchers of a particular gender, supposing the field and topic selection is a process of free choice depending on researchers' self-interest. Other studies attribute gender inequality to systemic biases against women[3], which act on publication outputs, citations, awarded grants, and collaborations[21-25][26-29]. Understanding what factors are and how they influence field or topic selection is still needed.

How interrelated and mutually shaping categories of region and gender served to compound knowledge inequalities remains a question to explore. Regional analysis should be introduced to solve this problem, but most research emphasizes the global gender or situation in a specific country[43]. Besides, previous research examines global disparities and gender inequalities in knowledge by bibliometric analysis with publications indexed in databases, yet unpublished knowledge carried by researchers is excluded. Online communication has made possible the rapid expansion of global science driven by human agents[44], providing a supplement to the traditional bibliometric perspective. More importantly, the scientific community-based question and answer gives a detailed description of researchers' knowledge status, while literature or professional survey data shows the research interest of researchers only. Researchers' attention to knowledge may come not only from their research interests but also from their needs in the process of scientific research. The question of researchers from the Q&A community can provide more information than scientific literature, thus providing a new perspective to catch sight of the relationship between regional and gender identities, fields, and topics.

Studies have demonstrated that under-represented gender groups produce high scientific novelty[45]. It is of scientific and social benefit to increase diversity and realize equality in global science[46] by first making out its manifestation. Therefore, the goal of this research is to analyze gender disparities within fields and within topics, looking for insights into region belongings and topic-based factors affecting underlying their knowledge requirement, at the same time describing field and topic selection and the relation between them in the Q&A scientific community. It extends investigations of gender disparities in science by providing a macro-level view of the phenomenon that accounts for the intersection of region, gender, field, and micro-level on the topic. The focus of knowledge carried by researchers in scientific community-based Q&A helps to reveal the distribution of researchers' knowledge requirements, which may behave in a different pattern with their research interest. The question and answer together show the knowledge status of region and gender groups, drawing the outline of their role in scientific discussion. This paper attempts to reveal the correlation between researchers' region and gender and their knowledge status in field and topic, with implications for better-achieving equality in science.

**Materials and Methods**

We collected data from the question part of ResearchGate by searching all the topics on the website from the year 2008 to May 2022 in about six months, which consisted of 460919 questions, 2760613 answers, and 31879 topics. The questions and answers are produced by 561362 distinct researchers. The first and last names of researchers are used to infer region and gender. The gender inferring algorithm is from Ferhat Elmas[47], which uses underlying data to cover the vast majority of first names in different countries. Gender is considered in a binary way as either 'male' or 'female,' as other genders can not be simply assigned, and data with unknown gender are excluded.

Ethnea is a collection of more than 9 million author name instances that are tagged one of the 26 ethnicity classes based on the name's association with national-level geo-locations [48]. To capture the gap of the region, we classify the ethnicity classes into eight categories according to geographical grouping, including 'East Asia,' 'Southeast Asia and the Pacific,' 'Sub-Saharan Africa,' 'North America and Western Europe,'' Arab States,' 'Central and Eastern Europe,' 'Latin America and the Caribbean' and 'South and West Asia.' By then, we get a table that summarizes the frequencies and ratios of region classes assigned by Ethnea to author name instances (SI Appendix, Dataset S1 and S2). The method raised by [49] is adopted to categorize the region of researchers, which is tested to be an empirical critique of name-based inference that minimizes bias.

Using first and last names, we compute each researcher's associated probability to each regional group instead of assigning the most probable group. We use these probabilities to compute weighted aggregates, in which each researcher contributes each group's aggregate as a function of the racial group with its name. That means, when computing the average mentions of topics by region, we assign the mentions of a topic fractionally to each regional group according to the corresponding distribution. In other words, we do not assign authors to a unique regional category. There are names whose last names in our ResearchGate data did not appear in the Ethnea data, names whose first names did not appear in the Ethnea data, and names whose last name or first name neither appeared in the Ethnea data. For those both first name and last name that do not appear in the Ethea data, each point of them is assigned as the mean distribution in the subset of researchers in the analysis. (For distribution of the researchers by region and gender in ResearchGate, see SI Appendix, Fig. S1)

Topics are defined according to the labels of ResearchGate with a multilevel hierarchy. The result contains 2265 first-level labels, and we manually classified each of them into six categories. The field classification of topics is based on the research areas of the Web of Science, in which Research areas are classified into five broad categories: 'Art & Humanities',' Life Science & Biomedicine,' 'Physical Science,' 'Social Science' ,and 'Technology'. Those topics which are not academic are concluded to be 'Others' and are not included in this research.

**Result**

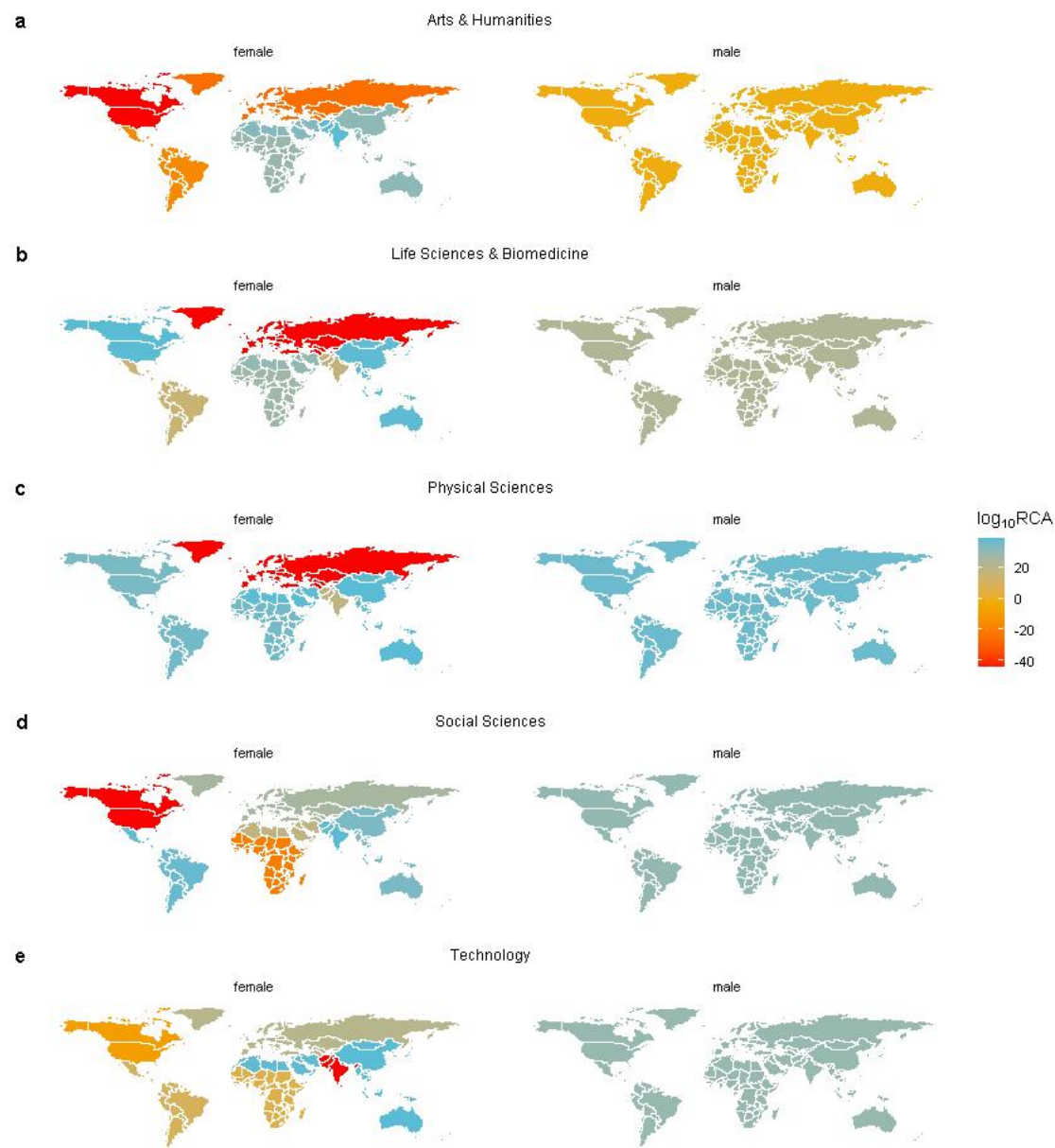

**Fig. 1.** Distribution of region and gender of researchers in question by field. The average number of questions and answers per researcher in each region and gender group. The average proportion of researchers in five fields normalized by region and gender group.

Men provide more answers compared with women in the same region (Fig. 1). Although the questions raised by men are slightly more than those of women, the gap between the answer and question number of men is wider than that of women. Women prefer to raise questions rather than provide answers, reflecting that women do not have as much knowledge compared with men. Such a gap indicates that the knowledge requirement of women is an issue worth exploring.

Fig. 1 then shows the comparison of distributions of researcher knowledge requirements in fields and varieties in regions and genders. In East Asia, North America and Western Europe, Central and Eastern Europe, Latin America and the Caribbean, and South and West Asia, women require more knowledge in Life Sciences & Biomedicine, and Social Sciences while men in Arts & Humanities, Physical Sciences and Technology. In North America, Western Europe, and Central and Eastern Europe, men require less knowledge of Physical Sciences and Technology. Women especially require knowledge in Social Sciences in North America and Western Europe and Life Sciences & Biomedicine in Latin America and the Caribbean. In Southeast Asia and the Pacific, women require more knowledge in Arts & Humanities, Life Sciences & Biomedicine, while men in Physical Sciences, Social Sciences, and Technology. There is a similar situation in Sub-Saharan Africa but in a more unbalanced distribution. Both men and women require more knowledge in Arts & Humanities and Social Sciences but less active in Physical Sciences and Technology. In the Arab States, women require more knowledge in Life Sciences & Biomedicine, and Physical Sciences. Overall, men tend to require more knowledge in Physical Sciences and Technology and women in Life Sciences & Biomedicine. Social Sciences is also a field in which, generally, women require more knowledge than men in regions except in Sub-Saharan Africa.

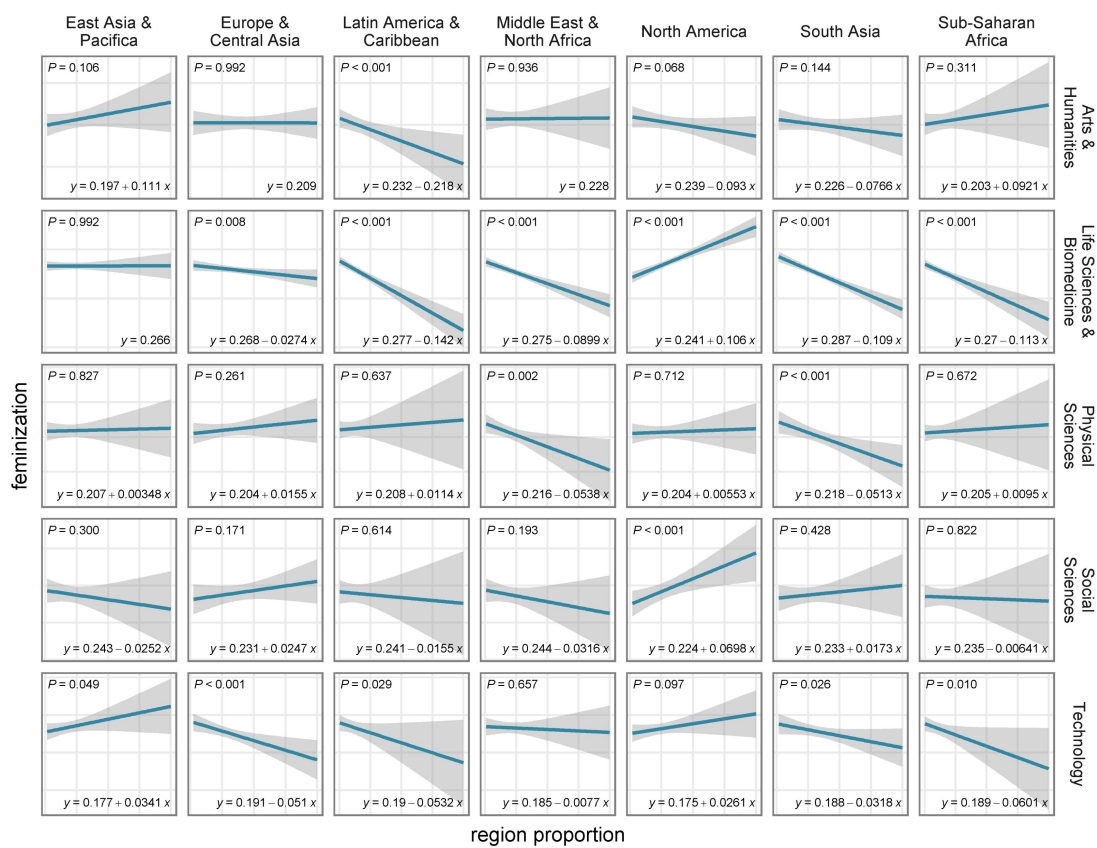

**Fig. 2.** The relationship between region and gender participation in question. The vertical axis shows the proportion of women, while the horizontal axis shows the regional proportion of researchers in each topic. To show the relationship clearer, topic points are omitted in the figure. The blue line shows the result of linear regression, together with the p-value in the upper left and the regression equation in the bottom.

To further reveal the knowledge requirement of women in the field from a micro-level of topics in a field and to better understand the relationship between region and gender participation, we analyze the region proportion and feminization of the topic using linear regression. Fig. 2 shows the fit curve and equation of linear regression. If women are equally distributed in topics of a discipline, the feminization (y value) should remain constant; if not so, unequal women distribution is then reflected. The negative regression coefficient means that with the increase of researchers in topics, the proportion of women taking part in the topics becomes less. It reflects that women are more concentrated on specific topics with fewer participators in the region, particularly with a high y-intercept. The non-negative regression coefficient represents that as more researchers participate in the topics, the proportion of women participation does not decrease, which means that women are actively taking part in topics in the entire discipline. In North America and Western Europe, the coefficient is significantly positive in Life Sciences & Biomedicine, Physical Sciences, Social Sciences, and Technology. There are also positive coefficients in Central and Eastern Europe in Life Sciences & Biomedicine and Latin America and the Caribbean in Life Sciences & Biomedicine and Physical Science. In East Asia, Arab States, and South and West Asia, the coefficient is significantly negative in Life Sciences & Biomedicine, Physical Sciences, Social Sciences, and Technology. There are also negative coefficients in Southeast Asia and the Pacific in Physical Sciences and Technology and in Sub-Saharan Africa in Life Sciences & Biomedicine, and Social Sciences. We find that in most regions, the regression coefficients remain positive or negative in different fields, showing that women's participation in topics is much related to region condition, except for Sub-Saharan Africa and Central and Eastern Europe.

We can see that in some regions of the world, women are still working on specific topics in scientific research. To further explore it, we select the question in the field of Life Sciences & Biomedicine to make a concrete analysis in Fig.3.

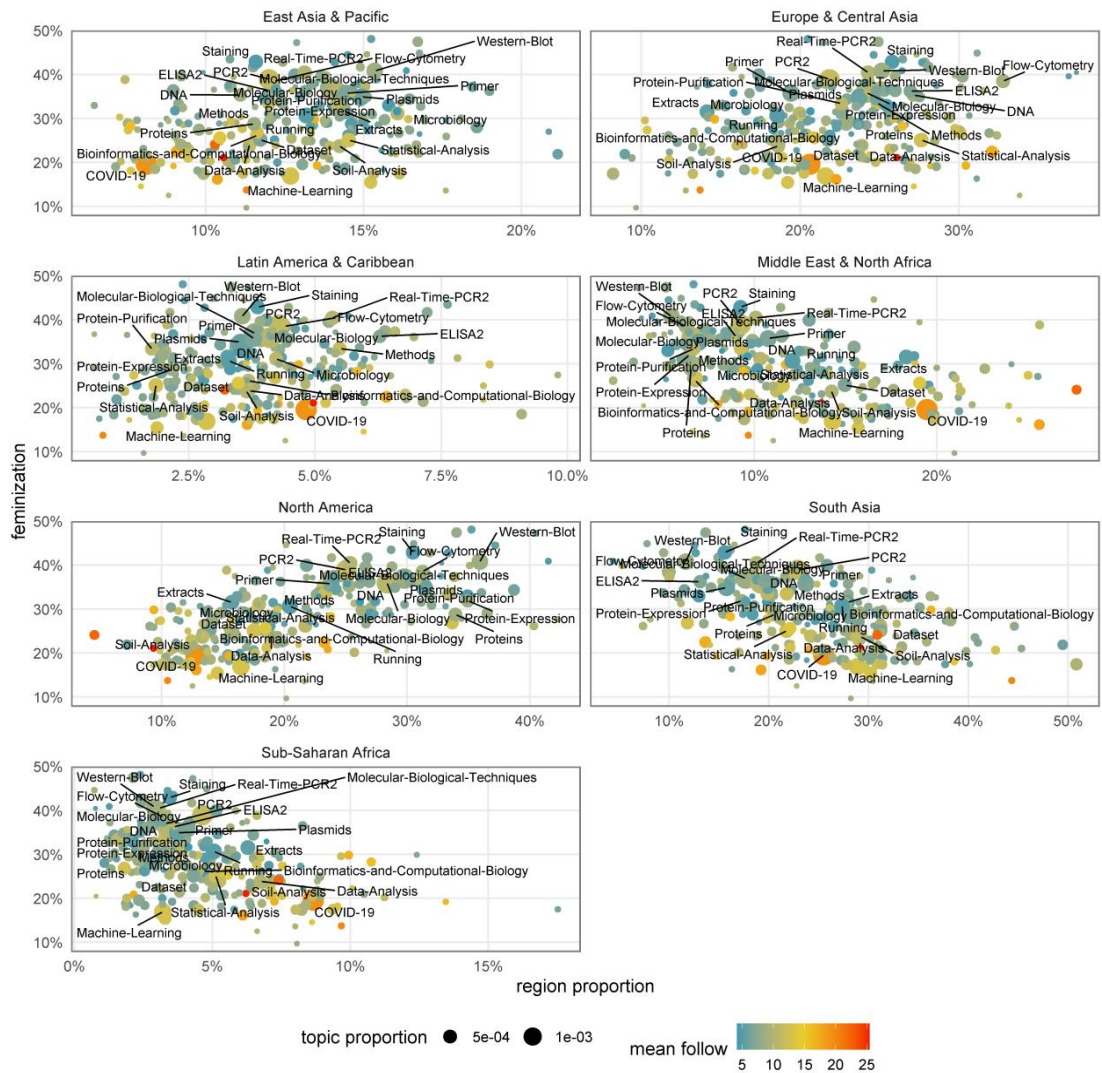

**Fig. 3.** Distribution of topic by region and gender participation in question. The vertical axis shows the proportion of women while the horizontal axis shows the region proportion in questioners of each topic. The color of the point shows the mean follow of a question on the topic. Two hundred fifty topics with the highest proportion are shown in the figure, and the top 20 topics are text-labeled.

The topics with a higher feminization in questioning in different regions are similar (Fig. 3). Staining (a method of imparting color to cells, tissues, or microscopic components), ELISA2 (a type of immunoassay that is commonly used to quantify levels of a specific target within a sample), Western-Blot (a laboratory method used to detect specific protein molecules from among a mixture of proteins) and Flow-Cytometry (a lab test used to analyze characteristics of cells or particles) are the topics of high feminization in all regions. And they are all about laboratory techniques. On the contrary, general topics like COVID-19 and Analysis are of low feminization and are majorly participated by men. This reveals the division of labor between men and women during the process of Life Sciences & Biomedicines research. Women are more involved in experimental operations than general research topics.

Staining is a specific topic with a low mean follow, and COVID-19 is a general topic with a high mean follow. They are selected as representations of specific and general topics in the following analysis. Based on the question-and-answer relationship in each topic, directed and weighted networks using the questioner and respondent as nodes are constructed. In the networks, the direction of the relationship is from the respondent nodes to the questioner nodes, representing the knowledge flow. The times that a pair of questioners and respondents interact act as the weight of the edge. The characteristics and organization forms of the topic networks are shown in Fig. 4.

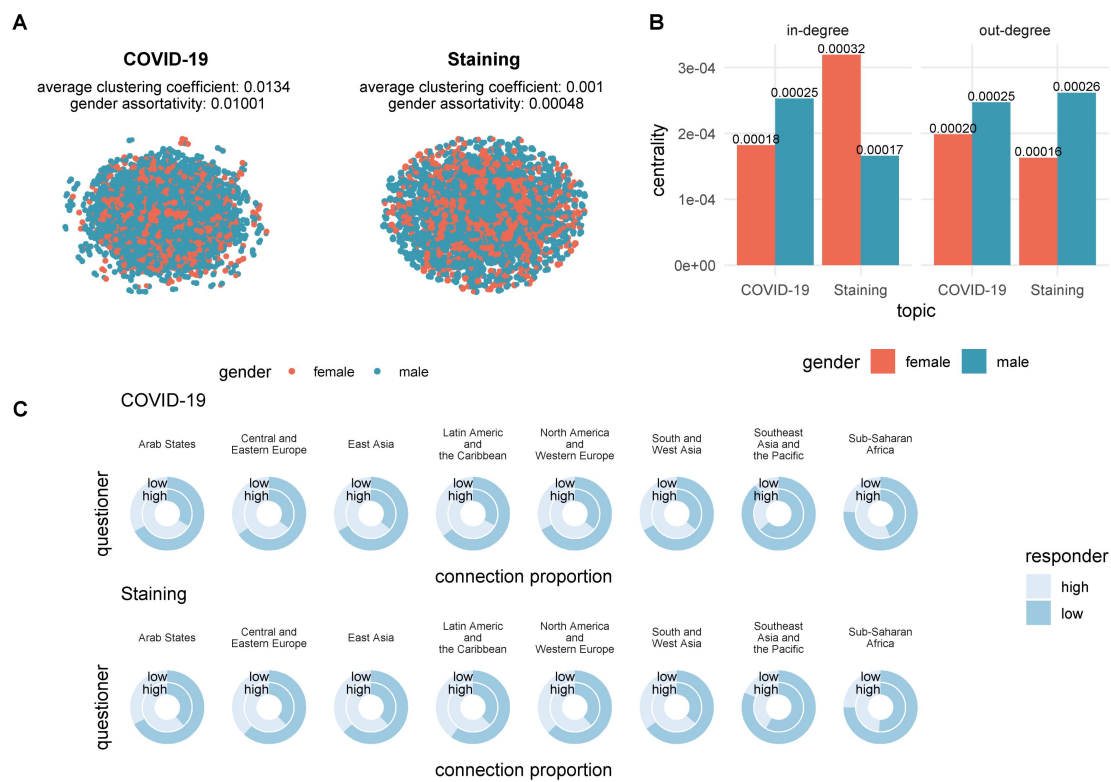

**Fig. 4.** Overall characteristics and organization forms of the topic networks. Network graph of the two topics with their average clustering coefficient (A). For better visualization effects, individual points in the network are not shown. The whole network can be seen in the supplementary material(SI Appendix, Fig. S). Average in-degree and out-degree centrality of female and male nodes (B). Connection proportion between questioner and responder of low and high probability (C). The low/high probability is divided by the median.

The clustering coefficient is a metric that measures how close a node is to forming a clique with its neighbors. Specifically, it refers to the degree to which adjacent points of a point are interconnected. The clustering coefficient of the COVID-19 network is larger than that of Staining (Fig. 4A), which shows that compared with a general discussed topic, the network of a specific topic is more fragmented with more little communities, while the network of a general topic is of better connectivity and forms larger communities. Knowledge flow in the latter one is not as good as in the former. Assortativity refers to the tendency of nodes to connect with other 'similar' nodes over 'dissimilar' nodes. Gender assortativity helps us to understand if researchers tend to connect to researchers of the same gender or not. Since the values of gender assortativity in both networks are greater than zero, both network is so-called assortative in gender (Fig. 4A). Yet the assortativity value of COVID-19 is higher than that of Staining, which shows that there is a more obvious tendency for researchers to answer questions raised by researchers with same gender about COVID-19 and to answer questions raised by researchers with different gender about Staining.

We compared women's and men's knowledge mastery of these two topics by computing the in and out the centrality of these two types of nodes in the Q&A network (Fig. 4B). An out relationship represents the answer to a question, and an in relationship represents receiving an answer. The more out relationships for a node, the more answers it can provide, and the higher degree of knowledge mastery a man or woman has. Comparatively, the more in the relationship for a node, the more answers it received, and the higher degree of knowledge lack a man or woman has. In the network of COVID-19, both in-degree and out-degree centrality of men are greater than women, while in the network of Staining, men are greater in out-degree centrality, and women are greater in in-degree centrality. That is, women receive more but provide fewer answers about the Staining topic. Men dominate the discussion, and women take less part in general topics, while women lack knowledge compared with men in specific topics.

Since the region attribute of the node is a fraction of probability other than an assigned value, it is difficult to calculate its assortativity in the usual way. In this research, it is represented by the connection proportion between nodes with different levels of probability in each region (Fig. 4C). The proportion of questions raised by low-probability nodes answered by low-probability nodes is higher than that answered by high-probability nodes. The same situation occurs when turning to the questions raised by high-probability nodes. That means researchers tend to answer the questions raised by researchers of the same region in the Q&A progress.

**Discussion**

Inequalities in science have been for a long time, and many related phenomena have been found. It is a complex system constituted of many parts and is influenced by many factors. Interventions in gender inequality need knowledge of its expression. Although gender inequality is ubiquitous in global science, its presentation varies in different regions and fields. The regional-specific characteristics and common global patterns of gender inequalities in science are then an issue worth exploring.

Our results show that the knowledge requirement of men and women in science varies from both the region and the field and is in line with their participation in published research, showing that their knowledge requirement is basically out of practical need. Considering economic factors, in middle and lower-middle developing regions, researchers tend to take part in natural and physical science. In highly developed regions like North America and Western Europe, there are more researchers participating in Arts & Humanities and Social Sciences, which represents a more balanced development between disciplines[44, 50, 51]. However, the same situation also occurs in developing regions like Sub-Sahara Africa, which may be because they do not have enough resources to support their research in natural and physical science. In the same field, the topics that researchers from different regions focus on are also various, which is likely due to the accumulated knowledge strength of regions. Besides, gender differences also affect researchers' knowledge requirements in fields. Although the geography of global science is evolving and emerging countries are rising, there exists a universal tendency that the Global North is more developed and balanced in its scientific portfolio[44, 50, 51], and so does the knowledge status of researchers. The disparities lie not only in the total science output that the Global North (established economies like Europe and the USA) far exceeds the Global South (emerging economies)[50] but also in the fields. Low-income countries tend to be more confined to natural research; middle countries extend towards the physical disciplines, and high-income countries tend to occupy both natural and societal research, as well as physical, suggesting balanced exploration[51]. The difference between men and women is more obvious: men usually need more knowledge in Physical Sciences and Technology, while women need more knowledge in Life Sciences & Biomedicines, and Social Sciences.

How women require knowledge among topics in a field also varies among regions. In regions with high economic development and more scientific accumulation, like North America and Western Europe, women require knowledge in topics all around the field, showing broad participation and wide interest. While in those regions with middle or low economic development and less scientific accumulation, their knowledge required topics are relatively limited to certain topics, which is not in line with a supposed equal distribution, giving a further description of the relationship between scientific participation and region related with inequalities. Such a universal tendency of women to work for certain topics may be a result of different gender stereotypes. With that, the development of the region's economy and science may lead to a change in researchers' gender perceptions, enabling women to participate widely in all topics in the field.

Women's field and topic distribution are generally influenced by regions, yet specific topics they focus on in a field are of common characteristics globally. We find that experimental operation is the kind of specific topic that women are anomalously requiring for knowledge by taking the field of Life Sciences & Biomedicines as an example. Moreover, it holds in all the regions, whatever their economic and scientific background. Meanwhile, it is also a universal tendency that women have less knowledge demand for more general topics, which means that their overall participation in general topics is not as good as men's to some extent. The result implicated that even though the development of modern science and the integration and intersection of fields make it difficult to say whether a field is purely experimental scientific research or purely theoretical scientific research, there is still a focus and division of labor in specific research work[52]. Such a division of work seems to be in line with the long-existing stereotype of women, who place more emphasis on experience rather than theory and on concrete rather than abstract[42].

Some previous studies recognize the gender difference both in the field and in a topic within the field as a display of research interest of men and women, attributing it to their different attitudes and ambition levels. These statements are at least partial. Supposing the above internal factors as the causes, then the knowledge status of women in their interested topics should be better, or at least should not be worse than that of men. However, by comparing women's and men's knowledge status using "Staining " as an example of a women-specific topic and "COVID-19" as an example of a general topic, we have observed that women can not meet their knowledge requirement, especially in the kind of specific topics that they are involved, telling that the inequality is of systemic biases against women including the influence of external factors[3]. In particular, such a difference in knowledge requirement distribution and related knowledge status can be a result of the labor division of women in scientific research.

The results reveal more. The knowledge status of women is worse than that of men, which is in line with the current situation that male authors were associated with greater scientific quality, particularly if the topic was male-typed[53]. Also, the online connection mode of the staining network seems more likely than man-woman collaboration in real-world research: the majority of male scientists collaborate with males, and most female scientists do not collaborate with females but with males[54]. These can be divided into several domains: resources, research outcomes, and monetary or non-monetary rewards like funding and prizes [55, 56]. Stereotypes and other people's views may also play a role in researchers' participation and research choice[57]. These factors together contribute to women's lack of early experience in some topics, as well as gaps in their self-efficacy[10], thus leading to their inferior knowledge status. Another implication is about the knowledge flow. Currently, the flow of knowledge in the scientific Q&A community is largely limited to the same regions and genders. However, it is an ideal pursuit of an open scientific community, which means that the knowledge flow network is heterogeneous from the perspective of degree as a characteristic of generally open social networks. However, the online communication network now is consistent with the collaboration network of scientists, in which researchers collaborate with same-gendered colleagues more often than expected[58]. The decreases in the diversity of scientific collaborations might negatively affect the progress of epistemic communities[59], and so does the online scientific community. Science communication online is like a nonstop academic conference for knowledge exchange[60]. Promoting the free flow of knowledge is significant for alleviating regional and gender knowledge inequality.

This paper has revealed the relation between region, gender, research fields, and research topics on gender inequalities in science. We infer that the economics and scientific accumulation of region decide their difference in field and gender participation, yet leave detailed research for verification. Quantitative correlation analysis, including region economics, scientific input, and output with indicators like GDP, R&D investment, and scientific literature output, can be conducted in future research. Furthermore, causal modeling that considers these variables would help to understand the mechanism of how systematic inequalities are formed and mediated. There are also other factors about cultural and social background that can affect inequalities in science, including faculty hiring[28, 29], career choice[30, 31], academic publishing[25], etc., with their related bias or stereotype. These are left for further exploration in their relation to intersectional inequalities of region and gender in science. Besides, a follow-up study can seek to directly display labor division differences if gender labor division data is available.

 Our analysis suggests systemic inequalities in science reflected in terms of the distribution of women and men in fields and in topics within fields, together with their knowledge status. Particularly, the narrow distribution of topics in fields and the low knowledge status of women are not only manifestations of inequalities but also stimuli of inequalities at the same time, as they may prevent women from gaining more funding related to broad research topics and winning good opinions in peer review[61]. This leads to some strategies to reduce inequalities for institutions, individuals, and scientific community platforms. Scientific institutions should recognize the existence of knowledge gaps related to region and gender and promote women's ability to participate widely in the field. Immediate actions can be taken to increase investment in teaching and learning resources in specific topics (e.g., experimental operation), specifically for women. Funding agencies can allocate increased funding in topics that have been less participated by a region or gender group so as to improve their productivity and knowledge accumulation in the fields or in the topics[62]. Correspondingly, individual researchers should be more active in participating in unexplored topics and areas. The scientific Q&A communities may improve their recommendation algorithms to change the community structure, better satisfy researchers' knowledge demand, and promote scientific knowledge flow. Taken together, these activities will serve to reduce both the field and topic knowledge disparities at the intersection of region and gender, thereby promoting equity and knowledge development in science.

# References:

[1]. UNESCO, "Women in Science" (UIS Fact Sheet No. 60 | June 2020). 2020.

[2]. Sugimoto, C.R., et al., Global gender disparities in science. NATURE, 2013. 504(7479): p. 211-213.

[3]. De Kleijn, M., et al., The Researcher Journey Through a Gender Lens: An Examination of Research Participation, Career Progression and Perceptions Across the Globe (Elsevier, March 2020). 2020.

[4]. Whittington, K.B., A tie is a tie? Gender and network positioning in life science inventor collaboration. RESEARCH POLICY, 2018. 47(2): p. 511-526.

[5]. Abramo, G., C.A. D'Angelo and G. Murgia, Gender differences in research collaboration. JOURNAL OF INFORMETRICS, 2013. 7(4): p. 811-822.

[6]. Liu, F.Y., et al., Gender inequality and self-publication are common among academic editors. NATURE HUMAN BEHAVIOUR, 2023.

[7]. Cho, A.H., et al., Women are underrepresented on the editorial boards of journals in environmental biology and natural resource management. PEERJ, 2014. 2.

[8]. Parker, M. and E.W. Welch, Professional networks, science ability, and gender determinants of three types of leadership in academic science and engineering. LEADERSHIP QUARTERLY, 2013. 24(2): p. 332-348.

[9]. Ceci, S.J., et al., Women in Academic Science: A Changing Landscape. Psychological Science in the Public Interest A Journal of the American Psychological Society, 2014. 15(3): p. 75-141.

[10]. Cheryan, S., et al., Why are some STEM fields more gender balanced than others? Psychological bulletin, 2017. 143(1): p. 1-35.

[11]. Huang, J., et al., Historical comparison of gender inequality in scientific careers across countries and disciplines. The National Academy of Sciences, 2020. 117.

[12]. Clayton, J.A. and F.S. Collins, Policy: NIH to balance sex in cell and animal studies. Nature, 2014. 509(7500): p. 282-283.

[13]. Klein, S.L., et al., Sex inclusion in basic research drives discovery. Proceedings of the National Academy of Sciences, 2015. 112(17): p. 5257-5258.

[14]. Koning, R., S. Samila and J. Ferguson, Who do we invent for? Patents by women focus more on women's health, but few women get to invent. Science, 2021. 372(6548): p. 1345-1348.

[15]. Sugimoto, C.R., et al., Factors affecting sex-related reporting in medical research: a cross-disciplinary bibliometric analysis. The Lancet, 2019. 393(10171): p. 550-559.

[16]. Lordan, G. and J.S. Pischke, Does Rosie Like Riveting? Male and Female Occupational Choices. ECONOMICA, 2022. 89(353): p. 110-130.

[17]. Lippa, R.A., K. Preston and J. Penner, Women's Representation in 60 Occupations from 1972 to 2010: More Women in High-Status Jobs, Few Women in Things-Oriented Jobs. PLOS ONE, 2014. 9(5).

[18]. Su, R. and R. James, All STEM fields are not created equal: People and things interests explain gender disparities across STEM fields. Frontiers in Psychology, 2015. 6: p. 189.

[19]. Bakshi-Hamm, P. and A. Hamm, Knowledge Production: Analysing Gender- and

Country-Dependent Factors in Research Topics through Term Communities. PUBLICATIONS, 2022. 10(4).

[20]. Kozlowski, D., et al., Intersectional inequalities in science. PROCEEDINGS OF THE NATIONAL ACADEMY OF SCIENCES OF THE UNITED STATES OF AMERICA, 2022. 119(2).

[21]. Beaudry, C. and V. Lariviere, Which gender gap? Factors affecting researchers' scientific impact in science and medicine. Research Policy, 2016. 45(9): p. 1790-1817.

[22]. Campbell, S.E. and S. Daniel, The Productivity Puzzle in Invasion Science: Declining but Persisting Gender Imbalances in Research Performance. BioScience, 2022(12): p. 12.

[23]. Fox, C.W., C.S. Burns and J.A. Meyer, Editor and reviewer gender influence the peer review process but not peer review outcomes at an ecology journal. Functional Ecology, 2015.

[24]. Jadidi, M., et al., Gender Disparities in Science? Dropout, Productivity, Collaborations and Success of Male and Female Computer Scientists. Advances in Complex Systems, 2017.

[25]. Kern-Goldberger, A.R., et al., The impact of double-blind peer review on gender bias in scientific publishing: a systematic review. AMERICAN JOURNAL OF OBSTETRICS AND GYNECOLOGY, 2022. 227(1): p. 43-+.

[26]. Romy, V.D.L. and N. Ellemers, Gender contributes to personal research funding success in The Netherlands. Proceedings of the National Academy of Sciences, 2015: p. 201510159.

[27]. Kim, L., et al., Gendered knowledge in fields and academic careers. RESEARCH POLICY, 2022. 51(1).

[28]. Way, S.F., D.B. Larremore and A. Clauset, Gender, Productivity, and Prestige in Computer Science Faculty Hiring Networks. PROCEEDINGS OF THE 25TH INTERNATIONAL CONFERENCE ON WORLD WIDE WEB (WWW'16). 2016. 1169-1179.

[29]. Blair-Loy, M., et al., DIVERSITY Can rubrics combat gender bias in faculty hiring? SCIENCE, 2022. 377(6601): p. 35-+.

[30]. Cundiff, J.L., et al., Do gender-science stereotypes predict science identification and science career aspirations among undergraduate science majors? SOCIAL PSYCHOLOGY OF EDUCATION, 2013. 16(4): p. 541-554.

[31]. Stout, J.G., V.A. Grunberg and T.A. Ito, Gender Roles and Stereotypes about Science Careers Help Explain Women and Men's Science Pursuits. SEX ROLES, 2016. 75(9-10): p. 490-499.

[32]. Rivera, L.A., When Two Bodies Are (Not) a Problem: Gender and Relationship Status Discrimination in Academic Hiring. American Sociological Review, 2017.

[33]. Barth, J.M., et al., Untangling Life Goals and Occupational Stereotypes in Men's and Women's Career Interest. SEX ROLES, 2015. 73(11-12): p. 502-518.

[34]. Leslie, S., et al., Expectations of brilliance underlie gender distributions across academic disciplines. Science, 2015. 347(6219): p. 262-265.

[35]. Livingstone, D.N., Putting Science in Its Place: Geographies of Scientific Knowledge. 2010: University of Chicago Press.

[36]. Kozlowski, J., S. Radosevic and D. Ircha, HISTORY MATTERS: THE INHERITED DISCIPLINARY STRUCTURE OF THE POST-COMMUNIST SCIENCE IN COUNTRIES OF CENTRAL AND EASTERN EUROPE AND ITS RESTRUCTURING. Scientometrics, 1999. 45(1): p. 137-166.

[37]. Boschma, R., G. Heimeriks and P.A. Balland, Scientific knowledge dynamics and relatedness in biotech cities. Research Policy, 2014. 43(1): p. 107-114.

[38]. Hidalgo, C.A., et al., The Principle of Relatedness, in Unifying Themes in Complex Systems IX

Proceedings of the Ninth International Conference on Complex Systems. 2018. p. 451-457.

[39]. Chinazzi, M., et al., Mapping the physics research space: a machine learning approach. EPJ DATA SCIENCE, 2019. 8(1).

[40]. Lee, L.C., et al., Research output and economic output: a Granger causality test. Scientometrics, 2011.

[41]. Stauvermann, et al., Exploring the link between research and economic growth: an empirical study of China and USA. Quality & Quantity: International Journal of Methodology, 2016.

[42]. Thelwall, M., et al., Gender differences in research areas, methods and topics: Can people and thing orientations explain the results? JOURNAL OF INFORMETRICS, 2019. 13(1): p. 149-169.

[43]. Thelwall, M., et al., Gender disparities in UK research publishing: Differences between fields, methods and topics. PROFESIONAL DE LA INFORMACION, 2020. 29(4).

[44]. Marginson, S., What drives global science? The four competing narratives. Studies in Higher Education, 2021(1): p. 1-19.

[45]. Hofstra, B., et al., The Diversity-Innovation Paradox in Science. PROCEEDINGS OF THE NATIONAL ACADEMY OF SCIENCES OF THE UNITED STATES OF AMERICA, 2020. 117(17): p. 9284-9291.

[46]. Papaioannou, T., Technological innovation, global justice and politics of development. PROGRESS IN DEVELOPMENT STUDIES, 2011. 11(4): p. 321-338.

[47]. SexMachine · PyPI. 2013.

[48]. Genni + Ethnea for the Author-ity 2009 dataset. 2018, University of Illinois at Urbana-Champaign.

[49]. Kozlowski, D., et al., Avoiding bias when inferring race using name-based approaches. 2021.

[50]. Gui, Q.C., et al., The changing geography of global science. ENVIRONMENT AND PLANNING A-ECONOMY AND SPACE, 2019. 51(8): p. 1615-1617.

[51]. Miao, L.L., et al., The latent structure of global scientific development. NATURE HUMAN BEHAVIOUR, 2022. 6(9): p. 1206-+.

[52]. Lin, S., Gender Study on the Status and Role of Women in Scientific Experiments. Journal of Chinese Women's Studies, 2011. 0(3): p. 48-53.

[53]. Knobloch-Westerwick, S., C.J. Glynn and M. Huge, The Matilda Effect in Science Communication An Experiment on Gender Bias in Publication Quality Perceptions and Collaboration Interest. Science Communication Linking Theory & Practice, 2013. 35(5): p. 603-625.

[54]. Kwiek, M. and W. Roszka, Gender-based homophily in research: A large-scale study of man-woman collaboration. Journal of Informetrics, 2021. 15(3): p. 101171.

[55]. Xie, Y., "Undemocracy": inequalities in science. Science, 2014. 344(6186): p. 809-810.

[56]. Meho, L.I., The gender gap in highly prestigious international research awards, 2001-2020. QUANTITATIVE SCIENCE STUDIES, 2021. 2(3): p. 976-989.

[57]. Penner, A.M., Gender inequality in science. Science, 2015. 347(6219): p. 234-235.

[58]. Holman, L. and C. Morandin, Researchers collaborate with same-gendered colleagues more often than expected across the life sciences. PLoS ONE, 2019. 14(4): p. e0216128-.

[59]. Rubin, H. and C. O'Connor, Discrimination and Collaboration in Science. Philosophy of ence, 2017. 85(3).

[60]. Foell, J., Social media science communication is a nonstop academic conference for all. NATURE HUMAN BEHAVIOUR, 2021. 5(7): p. 812-812.

[61]. Andersson, E.R., C.E. Hagberg and S. Haegg, Gender Bias Impacts Top-Merited Candidates.

Frontiers in Research Metrics and Analytics, 2021.

[62]. Jacob, B.A. and L. Lefgren, The impact of research grant funding on scientific productivity. JOURNAL OF PUBLIC ECONOMICS, 2011. 95(9-10): p. 1168-1177.